\documentclass[12pt, onecolumn, draftcls]{IEEEtran}

\usepackage{epsf}
\usepackage{graphics}
\usepackage{times}
\usepackage{graphicx}
\usepackage{amsmath}
\usepackage{amssymb}
\usepackage{epsfig}
\usepackage{subfigure}
\usepackage{array}
\usepackage{float}
\usepackage{bbm}

\newcommand{\E}{\mathbb{E}}
\newcommand{\Var}{\mathrm{Var}}
\newcommand{\Cov}{\mathrm{Cov}}
\newcommand{\CM}{\mathrm{CM}}
\newcommand{\dB}{\mathrm{dB}}
\newcommand{\RCM}{\mathrm{RCM}}

\newcommand{\Pav}{P_{\mathrm{av}}}

\begin{document}

\title{On the Properties of Cubic Metric\\ for OFDM Signals}

\author{Kee-Hoon Kim, Jong-Seon No, \textit{Fellow, IEEE}, and Dong-Joon Shin, \textit{Senior Member, IEEE}
\thanks{K.-H. Kim and J.-S. No are with the Department of Electrical and Computer Engineering, INMC, Seoul
National University, Seoul, 151-744, Korea (email: kkh@ccl.snu.ac.kr, jsno@snu.ac.kr).}
\thanks{D.-J. Shin is with the Department of Electronic Engineering, Hanyang University, Seoul, 133-791, Korea (email:
djshin@hanyang.ac.kr).}
}
\maketitle
\begin{abstract}
As a metric for amplitude fluctuation of orthogonal frequency division multiplexing (OFDM) signal, cubic metric (CM) has received an increasing attention because it is more closely related to the distortion induced by nonlinear devices than the well-known peak-to-average power ratio (PAPR). In this paper, the properties of CM of OFDM signal is investigated. First, asymptotic distribution of CM is derived. Second, it is verified that 1.7 times oversampling rate is good enough to capture the CM of continuous OFDM signals in terms of mean square error, which is also practically meaningful because the fast Fourier transform size is typically 1.7 times larger than the nominal bandwidth in the long-term evolution (LTE) of cellular communication systems.
\end{abstract}


\section{Introduction}
Orthogonal frequency division multiplexing (OFDM) is an attractive multicarrier modulation technique for broadband wireless access systems due to its strong immunity to multipath fading and high spectral efficiency. However, OFDM signals suffer from high amplitude fluctuation which causes performance degradation due to nonlinear devices. A well-known metric for amplitude fluctuation of OFDM signal is peak-to-average power ratio (PAPR). Many research efforts have been carried out to find efficient PAPR reduction techniques \cite{Lim}--\cite{TWC3}. Also, the distribution of the PAPR of continuous OFDM signal was derived \cite{Ochiai} and it is widely accepted that four times oversampling is enough to capture the PAPR of continuous OFDM signals \cite{Sharif}.

Another metric for amplitude fluctuation of OFDM signals has been considered \cite{Skrzypczak}\cite{Behravan}, which is known as cubic metric (CM) \cite{3GPP}. Studies on PAPR and CM suggest that, except for large power backoff, CM is more closely related to the amount of distortion induced by a nonlinear power amplifier than PAPR \cite{Behravan}. Moreover, after analyzing certain OFDM-type signals that are considered to meet the goal of the long-term evolution (LTE), it was shown in the 3GPP that CM predicts amplifier power de-rating more accurately than PAPR \cite{3GPP2}.

Thus, recent research to reduce the CM for the LTE systems has been carried out \cite{Zhu}--\cite{add2}. For example, in \cite{Zhu}, the clipping and filtering method to reduce the CM instead of PAPR is proposed, where the descent clipper different to the conventional clipper is designed. In \cite{Deumal}, the tone reservation method is proposed, where minimizing the CM value is set to the objective function of convex optimization.

A great deal of the literature has been devoted to research on the properties of the PAPR metric as in \cite{Ochiai}\cite{Sharif}. On the contrary, such analysis on the CM has not been done before. Thus, it is worth revealing more about the behavior of this metric. In this paper, an asymptotic probability distribution of CM for continuous OFDM signals is derived. Also, sufficient oversampling rate for capturing the CM of continuous OFDM signals is obtained.

\section{Cubic Metric}
The CM of OFDM signals is defined as \cite{3GPP2}
\begin{equation*}
\CM\big|_{\dB} \triangleq \frac{ \RCM\big|_{\dB}-\RCM_{\mathrm{ref}}\big|_{\dB}}{K}
\end{equation*}
where RCM is the raw CM defined by
\begin{equation}\label{eq:RCMdB}
\RCM[s(t)]\big|_{\dB} \triangleq 20\log \bigg[\mathrm{rms}\bigg[\bigg(\frac{|s(t)|}{\mathrm{rms}[s(t)]}\bigg)^3\bigg]\bigg]
\end{equation}
for a continuous OFDM signal $s(t)$ and both $\RCM_{\mathrm{ref}}\big|_{\dB}$ and $K$ are determined according to the considered OFDM systems \cite{3GPP2}. As an example, in the downlink of LTE, $\RCM_{\mathrm{ref}}\big|_{\dB} = 1.52~\dB$ and $K = 1.56$ are used. Thus, from (\ref{eq:RCMdB}), we are only interested in
\begin{equation*}
\RCM[s(t)] = \sqrt{\E\bigg[\bigg(\frac{|s(t)|}{\sqrt{\Pav}}\bigg)^6\bigg]}
\end{equation*}
where $\Pav$ is the average power of the continuous OFDM signal $s(t)$.
To simplify analysis, we will use the square of it as
\begin{equation*}
\xi \triangleq (\RCM[s(t)])^2 = \E\bigg[\bigg(\frac{|s(t)|}{\sqrt{\Pav}}\bigg)^6\bigg].
\end{equation*}

In practice, instead of calculating the CM of continuous OFDM signals, we calculate the CM of discrete OFDM signals.
Let $s(t)$ be a continuous OFDM signal and its $L$ times oversampled OFDM signal sequence be represented as
\begin{equation*}
s_{n,L} \triangleq s(nT_s/LN),~~~~~~0\leq n \leq LN-1
\end{equation*}
where $T_s$ is the OFDM signal period, $N$ is the number of subcarriers, and $L$ is a real number larger than or equal to one.

Without loss of generality, the input symbols in frequency domain are assumed to be statistically independent, identically distributed (i.i.d.) random variables with zero mean, where the input symbol is the complex data of each subcarrier. Then the OFDM signal components in time domain are given by the sum of i.i.d. random variables.
Thus, from the central limit theorem (CLT), the magnitude of $s_{n,L}$ is Rayleigh distributed \cite{Ochiai}.
Therefore, if it is normalized as
\begin{equation*}
r(t) \triangleq |s(t)|/\sqrt{\Pav}
\end{equation*}
\begin{equation*}
r_{n,L} \triangleq |s_{n,L}|/\sqrt{\Pav},
\end{equation*}
the probability distribution functions (PDFs) of $r(t)$ and $r_{n,L}$ are given as
\begin{equation*}
f_{r(t)}(r) = f_{r_{n,L}}(r) = 2\;re^{-r^2}.
\end{equation*}
Finally, for the discrete OFDM signal sequence obtained by $L$ times oversampling, $\xi$ and $\RCM$ are expressed as
\begin{equation}\label{eq:xiL}
\xi_L = \frac{1}{LN}\sum_{n=0}^{LN-1} r_{n,L}^6
\end{equation}
and
\begin{equation*}
\RCM_L = \sqrt{\xi_L}.
\end{equation*}

\section{Properties of $r_{n,L}^6$}
In (\ref{eq:xiL}), $\xi_L$ is sample mean of $r_{n,L}^6$'s and thus in this section we investigate the properties of the random variable $r_{n,L}^6$. By the definition of Weibull distribution, the power transformation $w_{n,L} \triangleq r_{n,L}^6$ of the Rayleigh distributed random variable $r_{n,L}$ is known as Weibull distribution \cite{Sagias}.
The PDF and cumulative distribution function (CDF) of $w_{n,L}$ are given as
\begin{equation*}
f_{w_{n,L}}(w) = \frac{1}{3} w^{-\frac{2}{3}} \exp(-w^{\frac{1}{3}})
\end{equation*}
and
\begin{equation*}
F_{w_{n,L}}(w) = 1 - \exp(-w^{\frac{1}{3}}),
\end{equation*}
respectively.
The $k$th-order moment of $w_{n,L}$ is known as $\E[w_{n,L}^k] = \Gamma(1+3k)$, where $\Gamma(a) = (a-1)!$ is the Gamma function for an integer $x$ and $\E[\cdot]$ denotes expectation value. Then we have
\begin{equation}\label{eq:meanofw}
\E[w_{n,L}] = 3! = 6
\end{equation}
\begin{equation*}
\E[w_{n,L}^{2}] = 6! = 720
\end{equation*}
\begin{equation}\label{eq:variance}
\Var(w_{n,L}) = 684
\end{equation}
where $\Var(\cdot)$ denotes variance.

Now, we calculate the covariance of two random variables $w_{n,L}$ and $w_{n',L'}$, which will be denoted as $\Cov(w_{n,L},w_{n',L'})$.
For this, we first obtain it with the continuous time lag $\tau$, $\Cov(w(t),w(t+\tau))$, where $w(t) = r^6(t)$.
The joint moment of $w(t)$ and $w(t+\tau)$ is expressed as \cite{Sagias}
\begin{equation}\label{eq:jointmoment}
\E[w^p(t) w^q(t+\tau)] = (1-\rho_{\tau}^2)^{1+3p+3q} \Gamma(1+3p)\;\Gamma(1+3q)\;{}_2F_1(1+3p,1+3q;1;\rho_{\tau}^2)
\end{equation}
where $_2F_1(\cdot,\cdot;\cdot;\cdot)$ is the Gauss hypergeometric function and $\rho_{\tau} = \rho_{s(t),s(t+\tau)}$ is the Pearson's correlation coefficient between $s(t)$ and $s(t+\tau)$ \cite{Sagias}.
From (\ref{eq:meanofw}) and (\ref{eq:jointmoment}), we have
\begin{align*}
\Cov(w(t),w(t+\tau)) = \E[w(t) w(t+\tau)] - \E[w(t)]\E[w(t+\tau)] = 36 (9 \rho_{\tau}^2 + 9 \rho_{\tau}^4 + \rho_{\tau}^6).
\end{align*}

It can be assumed as in \cite{Ochiai} that the power spectrum of baseband OFDM signal has conjugate symmetry, which is valid because the power spectrum of baseband OFDM signal can be designed to have symmetry at the center of the bandwidth by giving proper frequency offset. This assumption guarantees the autocorrelation function of $s(t)$, $\E[s(t)s(t+\tau)^*]$, to be a real function. In any case, the frequency offset is immaterial to our analysis, because it does not change the magnitude of the envelope of OFDM signal. In this case, the normalized autocorrelation function $\rho_{\tau}$ is known as \cite{Ochiai}
\begin{equation*}
\rho_{\tau} = \frac{\sin(\pi N \tau /T_s)}{N\;\sin(\pi \tau /T_s)}
\end{equation*}
for $\tau \neq 0$ and clearly $\rho_{\tau}=1$ for $\tau = 0$. Note that $\rho_{\tau}=0$ when
\begin{equation*}
\tau = \pm\frac{T_s}{N},\pm\frac{2T_s}{N},\pm\frac{3T_s}{N},\cdots
\end{equation*}
which implies the well known fact that the elements of the Nyquist sampled OFDM signal sequence are mutually independent.

The time lag between two discrete samples $w_{n,L}$ and $w_{n',L'}$ is
\begin{equation*}
\tau = \frac{nT_s}{LN} - \frac{n'T_s}{L'N}.
\end{equation*}
Finally, the covariance of $w_{n,L}$ and $w_{n',L'}$ is given as
\begin{equation}\label{eq:cov}
\Cov(w_{n,L},w_{n',L'}) = 36 (9 \rho_{\tau}^2 + 9 \rho_{\tau}^4 + \rho_{\tau}^6)\bigg|_{\tau = \frac{nT_s}{LN} - \frac{n'T_s}{L'N}}.
\end{equation}

\section{Distribution of RCM}
In this section, we obtain the asymptotic distribution of $\RCM_L$ by investigating the distribution of $\xi_L$ first.
\subsection{Mean and Variance of $\xi_L$}
The mean of $\xi_L$ is clearly 6 from (\ref{eq:meanofw}) because $\xi_L$ is sample mean of $w_{n,L}$.
To find the variance of $\xi_L$, suppose that $LN$ is an odd integer. Even though $LN$ can be any real number larger than or equal to $N$, it is not difficult to show that discrepancy by the assumption is negligible. Since $s_{n,L}$ is a complex stationary Gaussian process, both $r_{n,L}$ and $w_{n,L}$ are also stationary random process. Therefore, the variance of $\xi_L$ becomes
\begin{equation}\label{eq:varofprocess}
\Var(\xi_L) = \sigma_L^2 = \frac{\Var(w_{n,L})}{LN} + \frac{2}{(LN)^2}\sum_{k=1}^{LN-1}(LN-k)\;\Cov(w_{0,L},w_{k,L}).
\end{equation}
We can separate the summation in (\ref{eq:varofprocess}) into two parts and change the index of variable as
\begin{equation}\label{eq:sigmaL}
\sigma_L^2 = \frac{\Var(w_{n,L})}{LN} + \frac{2}{(LN)^2}\bigg(\sum_{k=1}^{\frac{LN-1}{2}}(LN-k)\;\Cov(w_{0,L},w_{k,L})+\sum_{k=1}^{\frac{LN-1}{2}}k\;\Cov(w_{0,L},w_{LN-k,L})\bigg).
\end{equation}
Using (\ref{eq:variance}) and the fact that the covariance $\Cov(w_{0,L},w_{k,L})$ in (\ref{eq:cov}) is symmetric and periodic with the period $LN$, (\ref{eq:sigmaL}) is rewritten as
\begin{equation*}
\sigma_L^2 = \frac{684}{LN} + \frac{2}{LN}\sum_{k=1}^{\frac{LN-1}{2}}\;\Cov(w_{0,L},w_{k,L}).
\end{equation*}
Consider the following two extreme cases using (\ref{eq:cov}).

\subsubsection{For the Nyquist Sampling Rate}
We have
\begin{equation*}
\sigma_1^2 = \frac{684}{N}.
\end{equation*}
\subsubsection{For the Continuous OFDM Signal}
We have
\begin{align}\label{eq:int}
\sigma_{\infty}^2 &= \lim_{L\rightarrow \infty} \sigma_L^2 = \lim_{L\rightarrow \infty}\frac{2}{LN}\sum_{k=1}^{\frac{LN-1}{2}}\;\Cov(w_{0,L},w_{k,L})\nonumber\\
&=\frac{72}{N\pi}\lim_{L\rightarrow \infty}\sum_{k=1}^{\frac{LN-1}{2}}\;\bigg(9\bigg(\frac{\sin(\frac{k\pi}{L})}{N\sin(\frac{k\pi}{LN})}\bigg)^2 + 9 \bigg(\frac{\sin(\frac{k\pi}{L})}{N\sin(\frac{k\pi}{LN})}\bigg)^4 + \bigg(\frac{\sin(\frac{k\pi}{L})}{N\sin(\frac{k\pi}{LN})}\bigg)^6\bigg)\frac{\pi}{L}\nonumber\\
&= \frac{72}{N\pi} \int_{0}^{\frac{N\pi}{2}} 9\bigg(\frac{\sin(x)}{N\sin(\frac{x}{N})}\bigg)^2 + 9 \bigg(\frac{\sin(x)}{N\sin(\frac{x}{N})}\bigg)^4 + \bigg(\frac{\sin(x)}{N\sin(\frac{x}{N})}\bigg)^6~dx\\
&= \frac{36}{5N^5}+\frac{117}{N^3}+\frac{2799}{5N}\nonumber.
\end{align}
Unless $N$ is too small, it becomes approximately $\sigma_{\infty}^2 \approx 2799/5N$. The detailed derivation of the integration in (\ref{eq:int}) is explained in Appendix.

\subsection{Distribution of $\xi_L$}\label{sec:Distribution}
\subsubsection{The Nyquist Sampling Rate Case ($L=1$)}
In this case, $w_{n,1}$'s are \textit{i.i.d.} and thus $\xi_1$ is asymptotically Gaussian distributed due to CLT. That is,
\begin{equation*}
\xi_1 \overset{a.s.}{\sim} \mathcal{N}\bigg(6,\sigma_{1}^2 \bigg)
\end{equation*}
where \textit{a.s.} means that the random variable is asymptotically distributed.
\subsubsection{Natural Number Sampling Rate Case ($L$ is a natural number)}
It is easy to see that in this case, $\xi_L$ is asymptotically Gaussian distributed.
We divide $LN$ OFDM signal components $s_{n,L}$ into $L$ subsets $\mathcal{S}_0, \mathcal{S}_1, \cdots, \mathcal{S}_{L-1}$ in the interleaved pattern as
\begin{align*}
\mathcal{S}_q = \{s_{n,L}\;\big|\; n = Lp +q,~0\leq p\leq N-1\},~~~~~0\leq q\leq L-1.
\end{align*}
Since the components $s_{n,L}$ in the set $\mathcal{S}_q$ are mutually independent, CLT can be applied to each set. Then we have $L$ asymptotically Gaussian distributions that are correlated. The sum of correlated Gaussian random variables is still Gaussian.
Therefore, when $L$ is a natural number, we have
\begin{equation*}
\xi_{L} \overset{a.s.}{\sim} \mathcal{N}\bigg(6,\sigma_{L}^2 \bigg).
\end{equation*}

\subsubsection{Continuous OFDM Signal Case ($L \rightarrow \infty$)}
Clearly, the distribution of $\xi_{\infty}$ converges to some distribution as $L$ increases and we already checked that the distributions of $\xi_L$ for all natural number $L$ are asymptotically Gaussian distribution. Therefore, the distribution of $\xi_{\infty}$ is also Gaussian such that
\begin{equation*}
\xi_{\infty} \overset{a.s.}{\sim} \mathcal{N}\bigg(6,\sigma_{\infty}^2 \bigg).
\end{equation*}

\subsubsection{$L$ is a Real Number}
Under some assumptions, the above three cases can be integrated into one general result. The CLT can be applied to $m$-dependent random process \cite{Hoeffding}, where $m$-dependent means that two samples from a random process with the interval larger than $m$ have no statistical dependency. $w_{n,L}$ is an $m$-dependent random process from the fact that the correlation of the stationary random process $w_{n,L}$ may rapidly diminish as the interval exceeds $T_s/N$ \cite{Ochiai}.
Therefore, for any real number $L$ larger than one, $\xi_L$ can be considered as Gaussian distributed.

\subsection{Distribution of $\RCM_L\big|_{\dB}$}
For the Gaussian distributed $\xi_L$ with mean 6 and variance $\sigma_L^2$, complementary CDF (CCDF) of $\xi_L$ is given as
\begin{equation*}
P(\xi_L > a) = 1 - F_{\xi_L}(a) = \frac{1}{2}\bigg[1 - \mathrm{erf}\bigg(\frac{a - 6}{\sigma_L \sqrt{2}} \bigg) \bigg].
\end{equation*}
Thus, CCDF of $\RCM_L\big|_{\dB}$ is
\begin{align*}
P(\RCM_L \big|_{\dB} > a) &= P(\xi_L > 10^{\frac{a}{10}})\nonumber\\
&= \frac{1}{2}\bigg[1 - \mathrm{erf}\bigg(\frac{10^{\frac{a}{10}} - 6}{\sigma_L \sqrt{2}} \bigg) \bigg]
\end{align*}
where $\mathrm{erf}(a)=\frac{2}{\sqrt{\pi}}\int_{0}^{a} e^{-t^2}~dt$ is the error function.
\begin{figure}[H]
  \includegraphics[width=\linewidth]{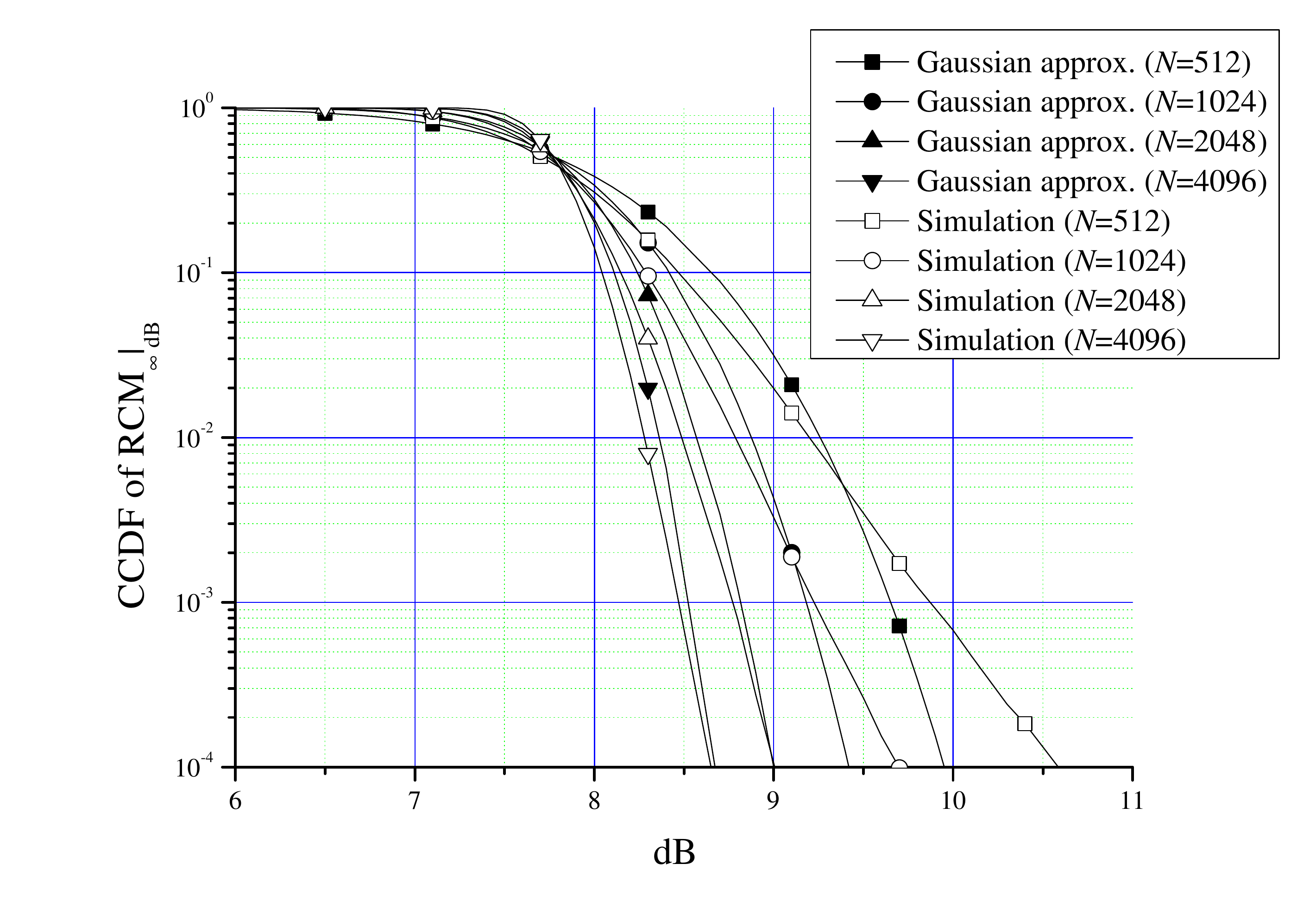}
  \caption{Comparison of simulated and analytical CCDFs of $\RCM_{\infty} \big|_{\dB}$.}
  \label{fig:continuous}
\end{figure}
Fig. \ref{fig:continuous} compares the simulated and analytical CCDFs of $\RCM_{\infty} \big|_{\dB}$. Though we have obtained the distribution of $\RCM_L\big|_{\dB}$ for general value of $L$, due to lack of space, we only present the comparison when $L\rightarrow \infty$, which is of importance practically. In the simulation, $L$ is set to 32, which is enough to represent the continuous OFDM signal, and 16-quadrature amplitude modulation is used. It is widely known that the metrics describing the envelope behavior of OFDM signals such as CM and PAPR do not depend on the modulation order. Note that analysis is based on Gaussian approximation in Section \ref{sec:Distribution}. Unfortunately, Weibull distribution is a kind of heavy-tailed distribution and thus the sum of Weibull random variables slowly converges to Gaussian distribution as $N$ increases but it shows good agreement when $N$ is large as shown in Fig. \ref{fig:continuous}.

\section{Oversampling Rate to Capture the CM of Continuous OFDM Signal}

Radically, the metrics such as PAPR and CM have to be calculated from continuous OFDM signals. However, it is impossible to handle with the continuous OFDM signal and instead one calculates the metrics from the oversampled discrete OFDM signal with sufficient oversampling rate.
In this case, the metrics calculated from the oversampled OFDM signal can be viewed as an estimator for the metrics of the continuous OFDM signal. For instance, in the case of the CM calculation, $\RCM_L$ is the estimator for true parameter $\RCM_{\infty}$. It is natural that a high sampling rate guarantees low estimation error but it also entails high complexity. Thus, finding a sufficient sampling rate is of great importance.

In \cite{Sharif}, upper bounds on the estimation error of PAPR according to $L$ is derived and it is proposed that four times oversampling rate is enough to capture the PAPR of continuous OFDM signals. However, in the case of CM, this approach is not useful. Differing from the PAPR, the estimator $\RCM_L$ rarely shows extremely large estimation error. Therefore, the total inspection approach is not suitable in the case of CM. This is due to the fact that the Weibull distribution is a kind of heavy-tailed distribution which means that its variance is quite large.
For instance, when the OFDM signal sequence is an impulse signal with $N=1024$, it has $\RCM_1 = 1024$ and $\RCM_{\infty} \simeq 759$. In this case, the square error $|\RCM_1 - \RCM_{\infty}|^2 \simeq 7\times 10^4$ is much larger than the mean square error (MSE) $\E[|\RCM_1 - \RCM_{\infty}|^2] \simeq 5\times 10^{-3}$. Thus, in this section we statistically approach this problem by deriving the MSE $\E[|\RCM_L - \RCM_{\infty}|^2]$ according to $L$.

\subsection{Joint PDF of $\xi_L$ and $\xi_{\infty}$}
To obtain the MSE $\E[|\RCM_L - \RCM_{\infty}|^2] = \E[|\sqrt{\xi_L}-\sqrt{\xi_{\infty}}|^2]$, first we find the joint PDF of $\xi_L$ and $\xi_{\infty}$ which is clearly bivariate Gaussian distribution from our investigation in Section \ref{sec:Distribution}. We already checked their mean values $\E[\xi_L]=\E[\xi_{\infty}]=6$.
Next, the correlation coefficient between $\xi_L$ and $\xi_{\infty}$ is given as
\begin{align*}
\rho_{\xi_L,\xi_{\infty}} &= \frac{\Cov(\xi_L,\xi_{\infty})}{\sigma_L\sigma_{\infty}}\\
&= \lim_{L'\rightarrow \infty}\frac{\sum_{n=0}^{LN-1}\sum_{n'=0}^{L' N-1}\Cov(w_{n,L},w_{n',L'})}{\sigma_L\sigma_{\infty}L L' N^2}\\
&= \lim_{L'\rightarrow \infty}\frac{\sum_{n=0}^{LN-1}\sum_{n'=0}^{L' N-1}36(9\rho_{\tau}^2 + 9\rho_{\tau}^4 + \rho_{\tau}^6)\big|_{\tau = \frac{nT_s}{LN} - \frac{n'T_s}{L'N}}}{\sigma_L\sigma_{\infty}L L' N^2}\\
&= \frac{36}{\sigma_L\sigma_{\infty}\pi L N^2}\cdot\sum_{n=0}^{LN-1}\int_{-\frac{n\pi}{L}}^{-\frac{n\pi}{L}+N\pi} 9\bigg(\frac{\sin(x)}{N\sin(\frac{x}{N})}\bigg)^2 + 9\bigg(\frac{\sin(x)}{N\sin(\frac{x}{N})}\bigg)^4 + \bigg(\frac{\sin(x)}{N\sin(\frac{x}{N})}\bigg)^6\:dx
\end{align*}
where the equation in the integration is periodic with the period $N\pi$. Thus, we have
\begin{align}\label{eq:rhotau}
\rho_{\xi_L,\xi_{\infty}}&= \frac{36}{\sigma_L\sigma_{\infty}\pi N}\cdot\int_{0}^{N\pi} 9\bigg(\frac{\sin(x)}{N\sin(\frac{x}{N})}\bigg)^2 + 9\bigg(\frac{\sin(x)}{N\sin(\frac{x}{N})}\bigg)^4 + \bigg(\frac{\sin(x)}{N\sin(\frac{x}{N})}\bigg)^6\:dx\nonumber\\
&=\frac{\sigma_{\infty}}{\sigma_L}.
\end{align}
In terms of estimation theory, $\rho_{\xi_L,\xi_{\infty}}=\sigma_{\infty}/\sigma_L$ implies that $\xi_L$ can be considered as an unbiased minimum MSE estimator of $\xi_{\infty}$. That is, orthogonality principle $\E[(\xi_L - \xi_{\infty})\;\xi_{\infty}] = 0$ and unbiased property $\E[\xi_L] = \E[\xi_{\infty}]$ are satisfied.

\subsection{MSE between $\xi_L$ and $\xi_{\infty}$}
Using the orthogonality principle, the MSE between $\xi_L$ and $\xi_{\infty}$ is  obtained as
\begin{equation}\label{eq:MSExiL}
\E[|\xi_L-\xi_{\infty}|^2] = \sigma_L^2 - \sigma_{\infty}^2.
\end{equation}

%\begin{figure}[H]
%  \includegraphics[width=\linewidth]{MSE_of_xi_compare.eps}
%  \caption{Comparison of $\MSE(\xi,L)$ of simulation results and analytical results.}
%\end{figure}
\begin{figure}[H]
  \center
  \includegraphics[width=0.85\linewidth]{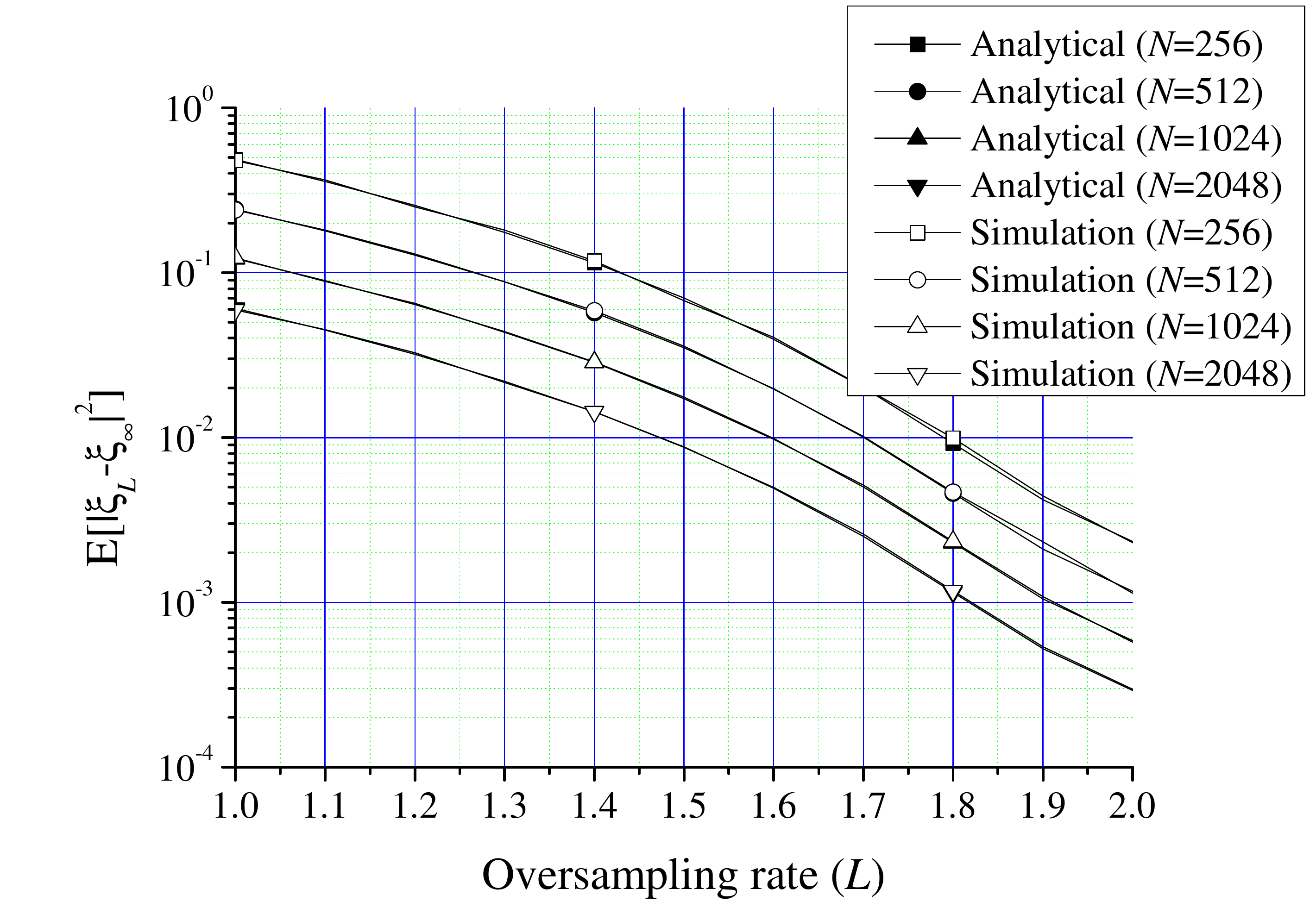}
  \caption{Comparison of simulated and analytical $\E[|\xi_L-\xi_{\infty}|^2]$ for various $L$.}
  \label{fig:MSE_of_xi}
\end{figure}
Fig. \ref{fig:MSE_of_xi} compares $\E[|\xi_L-\xi_{\infty}|^2]$ values obtained by simulation and analysis, which shows a good agreement. Analytical results are given by using (\ref{eq:MSExiL}) and simulation results are obtained by testing randomly generated $10^5$ OFDM signal sequences.

\subsection{MSE between $\RCM_L$ and $\RCM_{\infty}$}

The MSE between $\RCM_L$ and $\RCM_{\infty}$ can be obtained as
\begin{align*}
\E[|\RCM_L-\RCM_{\infty}|^2]
&= \E[|\sqrt{\xi_L}-\sqrt{\xi_{\infty}}|^2]\\
&= 12 - 2\cdot\E[\sqrt{\xi_L \xi_{\infty}}]\\
&\overset{a.s.}{\sim} 12 - 2 \int_{0}^{\infty}\int_{0}^{\infty} \sqrt{\xi_L \xi_{\infty}}\; f_{\xi_L,\xi_{\infty}}(\xi_1,\xi_2)~d\xi_1\;d\xi_2
\end{align*}
where $f_{\xi_L,\xi_{\infty}}(\xi_1,\xi_2)$ is the joint Gaussian PDF of $\xi_1$ and $\xi_2$ with the correlation coefficient $\rho_{\xi_L,\xi_{\infty}}$ in (\ref{eq:rhotau}).

\begin{figure}[H]
  \center
  \includegraphics[width=0.85\linewidth]{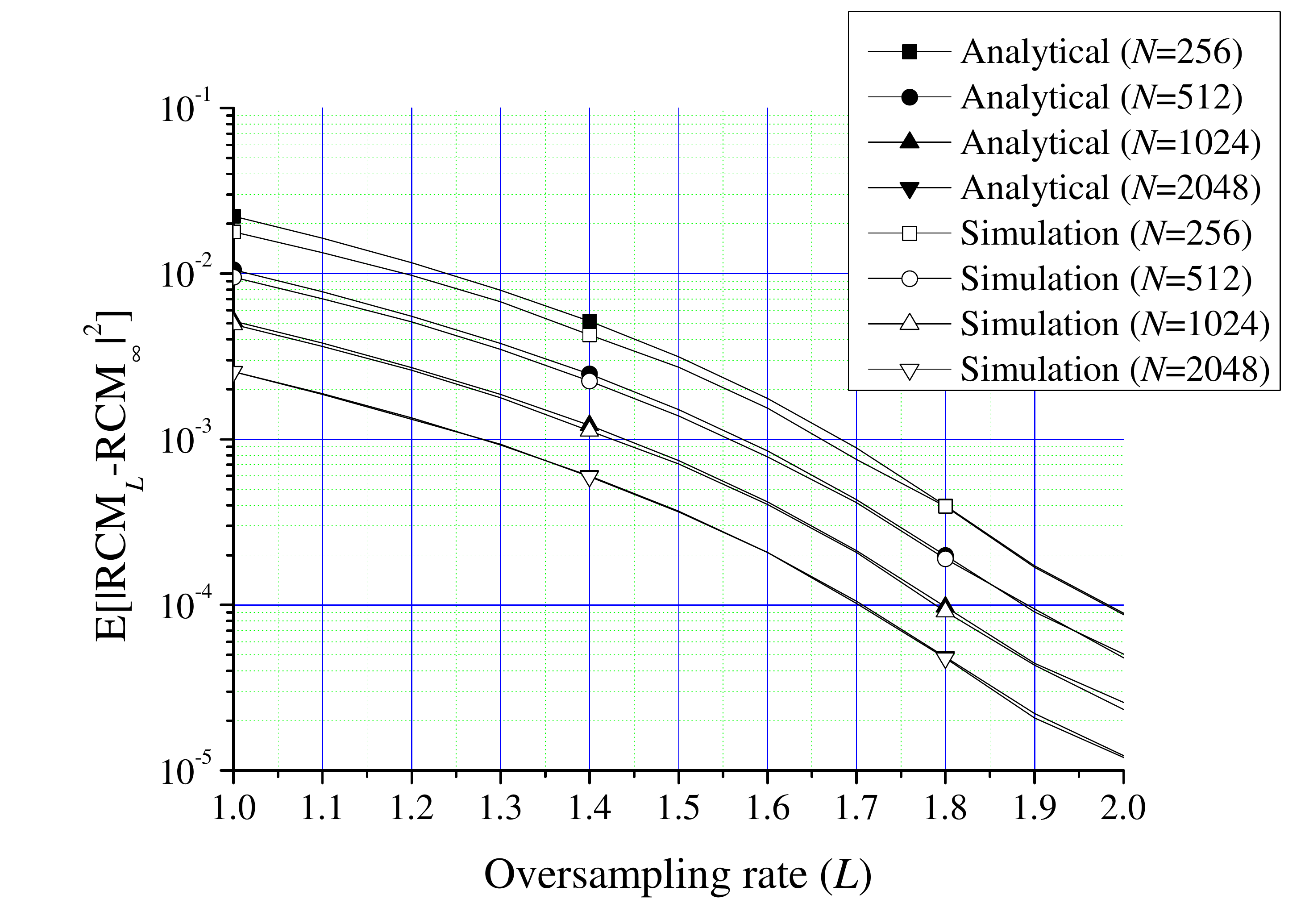}
  \caption{Comparison of simulated and analytical $\E[|\RCM_L-\RCM_{\infty}|^2]$ for various $L$ using Gaussian approximation.}
  \label{fig:MSE_of_RCM}
\end{figure}

Fig. \ref{fig:MSE_of_RCM} compares $\E[|\RCM_L-\RCM_{\infty}|^2]$ values obtained by simulation and analysis using Gaussian approximation.
From a practical viewpoint, MSE is usually normalized as $\E[|\RCM_L-\RCM_{\infty}|^2]/\E[|\RCM_{\infty}|^2] = \E[|\RCM_L-\RCM_{\infty}|^2]/6$.
Thus, one can conclude that 1.7 times oversampling gives the normalized MSE smaller than about $10^{-4}$ for practical value of $N \geq 256$, which is remarkable because the FFT size is typically 1.7 times larger than the nominal bandwidth for the LTE cellular communication systems \cite{3GPP3}.
%\begin{figure}[H]
%  \includegraphics[width=\linewidth]{MSE_of_RCM_compare.eps}
%  \caption{Comparison of $\MSE(\RCM,L)$ of simulation results and analytical results using Gaussian approximation.}
%\end{figure}

\section{Conclusion}
In this paper, the properties of CM are investigated. First, asymptotic distribution of the CM is obtained, which shows a good agreement to simulation results when the number of subcarriers is large. Second, the oversampling rate good enough to capture the CM of the continuous OFDM signal is investigated. We confirmed that 1.7 times oversampling rate is good enough for that purpose from a viewpoint of MSE, which is of great importance because the FFT size is typically 1.7 times larger than the nominal bandwidth for the LTE cellular communication systems.

\section*{Appendix}

\subsection*{Derivation of $\int_{0}^{\frac{N\pi}{2}} (\frac{\sin(x)}{N\sin(\frac{x}{N})})^r dx$ When $r=$2, 4, and 6}
It is known that there are several methods to derive $\int_{0}^{\frac{N\pi}{2}} (\frac{\sin(x)}{N\sin(\frac{x}{N})})^r dx$ when $r=2$. By expanding one of them, both cases of $r=4$ and $6$ are derived analogously as follows. Without loss of generality, by using $N=2M$ and change of variables, we have more simplified expression as
\begin{equation*}
\int_{0}^{\frac{N\pi}{2}} \bigg(\frac{\sin(x)}{N\sin(\frac{x}{N})}\bigg)^r dx = \frac{1}{2^r M^{r-1}}\int_{0}^{\pi} \bigg(\frac{\sin(2Mx)}{\sin(x)}\bigg)^r dx
\end{equation*}
where
\begin{equation}\label{eq:polyexp}
\bigg(\frac{\sin(2Mx)}{\sin(x)}\bigg)^r = 2^r(\cos(x) + \cos(3x) + \cdots + \cos((2M-1)x))^r.
\end{equation}
Since the integral is over $[0, \pi]$, only part of coefficients contribute to the integration. For example, when $r=2$, all the cross terms from the polynomial expansion in (\ref{eq:polyexp}) becomes zero after integration, i.e., $\int_0^{\pi} \cos(x)\cos(3x)dx=0$. After some manipulations, we have
\begin{align*}
&\bigg(\frac{\sin(2Mx)}{\sin(x)}\bigg)^2 = 2^2  \bigg(\frac{M}{2} + z_2(x)\bigg)\\
&\bigg(\frac{\sin(2Mx)}{\sin(x)}\bigg)^4 = 2^4  \bigg(\frac{1}{24}\bigg(M + 8M^3\bigg) + z_4(x)\bigg)\\
&\bigg(\frac{\sin(2Mx)}{\sin(x)}\bigg)^6 = 2^6  \bigg(\frac{1}{160}\bigg(M + 5 M^3 + 44 M^5\bigg) + z_6(x)\bigg)
\end{align*}
where $z_r(x)$ denotes all terms which become zero after integration over $[0, \pi]$. Finally, we have
\begin{align*}
&\int_{0}^{\frac{N\pi}{2}} \bigg(\frac{\sin(x)}{N\sin(\frac{x}{N})}\bigg)^2 dx = \frac{\pi}{2}\\
&\int_{0}^{\frac{N\pi}{2}} \bigg(\frac{\sin(x)}{N\sin(\frac{x}{N})}\bigg)^4 dx = \pi\bigg(\frac{1}{3} + \frac{1}{6N^2} \bigg)\\
&\int_{0}^{\frac{N\pi}{2}} \bigg(\frac{\sin(x)}{N\sin(\frac{x}{N})}\bigg)^6 dx = \pi\bigg(\frac{11}{40} + \frac{1}{8N^2} + \frac{1}{10N^4} \bigg).
\end{align*}


\begin{thebibliography}{99}

\bibitem{Lim}
D.-W. Lim, J.-S. No, C.-W. Lim, and H. Chung, ``A new SLM OFDM scheme with low complexity for PAPR reduction,'' \textit{IEEE Signal Process. Lett.}, vol. 12, no. 2, pp. 93--96, Feb. 2005.

\bibitem{Chen}
J.-C. Chen and C.-K. Wen, ``PAPR reduction of OFDM signals using cross-entropy-based tone injection schemes,'' \textit{IEEE Signal Process. Lett.}, vol. 17, no. 8, pp. 727--730, Aug. 2010.

\bibitem{Kim}
K.-H. Kim, H.-B. Jeon, J.-S. No, and D.-J. Shin, ``Low-complexity selected mapping scheme using cyclic-shifted inverse fast Fourier transform for peak-to-average power ratio reduction in orthogonal frequency division multiplexing systems,'' \textit{IET Commun.}, vol. 7, no. 8, pp. 774--782, May 2013.

\bibitem{TWC1}
S. Shu, D. Qu, L. Li, and T. Jiang, ``Invertible subset QC-LDPC codes for PAPR reduction of OFDM signals,'' \textit{IEEE Trans. Broadcast.}, vol. 61, no. 2, pp. 290--298, Feb. 2015.

\bibitem{TWC2}
M. Hu, Y. Li, W. Wang, and H. Zhang, ``A piecewise linear companding transform for PAPR reduction of OFDM signals with companding distortion mitigation,'' \textit{IEEE Trans. Broadcast.}, vol. 60, no. 3, pp. 532--539, Aug. 2014.

\bibitem{TWC3}
S.-J. Ku, ``Low-complexity PTS-based schemes for PAPR reduction in SFBC MIMO-OFDM systems,'' \textit{IEEE Trans. Broadcast.}, vol. 60, no. 4, pp. 650--658, Nov. 2014.

\bibitem{Ochiai}
H. Ochiai and H. Imai, ``On the distribution of the peak-to-average power ratio in OFDM signals,'' \textit{IEEE Trans. Commun.}, vol. 49, no. 2, pp. 282--289, Feb. 2001.

\bibitem{Sharif}
M. Sharif, M. Gharavi-Alkhansari, and B. H. Khalaj, ``On the peak-to-average power of OFDM signals based on oversampling,'' \textit{IEEE Trans. Commun.}, vol. 51, no. 1, pp. 72--78, Jan. 2003.

\bibitem{Skrzypczak}
A. Skrzypczak, P. Siohan, and J. Javaudin, ``Power spectral density and cubic metric for the OFDM/OQAM modulation,'' in \textit{Proc. IEEE
International Symp. Signal Process. Inf. Technol.}, Aug. 2006, pp. 846--850.

\bibitem{Behravan}
A. Behravan and T. Eriksson, ``Some statistical properties of multicarrier signals and related measures,'' in \textit{Proc. IEEE Veh. Technol. Conf. Spring}, vol. 4, May 2006, pp. 1854-1858.

\bibitem{3GPP}
3GPP TSG RAN WG1 and 3GPP TSG RAN WG4, TDocs R4-040367, R1-040522 and R1-040642, ``Comparison of PAR and cubic metric for
power de-rating,'' May 2004.

\bibitem{3GPP2}
3GPP TSG RAN WG1, TDoc R1-060023, ``Cubic metric in 3GPP-LTE,'' Jan. 2006.

\bibitem{Zhu}
X. Zhu, H. Hu, and Y. Tang, ``Descendent clipping and filtering for cubic metric reduction in OFDM systems,'' \textit{Electron. Lett.}, vol. 49, no. 9, pp. 599--600, Apr. 2013.

\bibitem{Deumal}
M. Deumal, A. Behravan, and J. L. Pijoan, ``On cubic metric reduction in OFDM systems by tone reservation,'' \textit{IEEE Trans. Commun.}, vol. 59, no. 6, pp. 1612--1620, Jun. 2011.

\bibitem{add1}
R. Y. Kim, Y. Y. Kim, A. A. Yazdi, S. Sorour, and S. Valaee, ``Joint reduction of peak-to-average power ratio, cubic metric, and block error rate in OFDM systems using network coding,'' \textit{IEEE Trans. Veh. Tech.}, vol. 60, no. 9, pp. 4363--4373, Oct. 2011.

\bibitem{add2}
J. G. Doblado, A. C. O. Oria, V. Baena-Lecuyer, P. Lopez, and D. Perez-Calderon, ``Cubic metric reduction for DCO-OFDM visible light communication systems,'' \textit{J. Lightwave Technol.}, vol. 33, no. 10, pp. 1971--1978, Feb. 2015.


\bibitem{Sagias} N. C. Sagias and G. K. Karagiannidis, ``Gaussian class multivariate Weibull distributions: Theory and applications in fading channels,'' \textit{IEEE Trans. Inf. Theory}, vol. 51, no. 10, pp. 3608--3619, Oct. 2005.

\bibitem{Hoeffding} W. Hoeffding and H. Robbins, ``The central limit theorem for dependent random variables,'' \textit{Duke Math. J.}, vol. 15, no. 3, pp. 773--780, 1948.

%\bibitem{Yacoub}
%M. D. Yacoub, ``The $\alpha-\mu$ distribution: a physical fading model for
%the Stacy distribution,'' \textit{IEEE Trans. Veh. Technol.}, vol. 56, no. 1, pp. 27--34, Jan. 2007.
%
%\bibitem{Reig}
%J. Reig, L. Rubio, and N. Cardona, ``Bivariate Nakagami-$m$ distribution with arbitrary fading parameters,'' \textit{Electron. Lett.}, vol. 38, no. 25, pp. 1715--1717, Dec. 2002.

\bibitem{3GPP3}
3GPP TSG RAN WG4, TS 36.104, ``Evolved universal terrestrial radio access (E-UTRA); Base station (BS) radio transmission and reception,'' May 2008.



\end{thebibliography}
\end{document}